\documentclass[onecolumn,superscriptaddress,groupedaddress]{revtex4-1}  
\usepackage{arxiv}
\usepackage{amsmath}
\usepackage{graphicx}  
\usepackage{dcolumn}   
\usepackage{bm}        
\usepackage{amssymb}   
\usepackage[normalem]{ulem}

\usepackage{tikz}
\usetikzlibrary{arrows, arrows.meta}
\usepackage{hyperref}
\hypersetup{
    colorlinks=true,
    linkcolor=blue,
    filecolor=blue,      
    urlcolor=black,
    citecolor=blue,
        }
\usepackage{subfigure}

\newcommand{\poubelle}[1]{}

\begin{document}
\title{Erythrocyte-erythrocyte aggregation dynamics  under shear flow}

\author{Mehdi Abbasi}
\email{mehdi.abbasi@univ-grenoble-alpes.fr}
\affiliation{Univ. Grenoble Alpes, CNRS, LIPhy, F-38000 Grenoble, France}
\author{Alexander Farutin}
\affiliation{Univ. Grenoble Alpes, CNRS, LIPhy, F-38000 Grenoble, France}
\author{Hamid Ez-Zahraouy} 
\affiliation{LaMCScI, Faculty of Sciences, Mohammed V University of Rabat, 1014 Morocco}
\author{Abdelilah Benyoussef}
\affiliation{LaMCScI, Faculty of Sciences, Mohammed V University of Rabat, 1014 Morocco}
\affiliation{Hassan II Academy of Science and Technology, Rabat 10220, Morocco}

\author{Chaouqi Misbah}
\email{chaouqi.misbah@univ-grenoble-alpes.fr}\affiliation{Univ. Grenoble Alpes, CNRS, LIPhy, F-38000 Grenoble, France}


\begin{abstract}
Red blood cells (RBCs) --erythrocytes-- suspended in plasma tend to aggregate and form rouleaux.  During aggregation the first stage consists in the formation  of RBC doublets [Blood cells, molecules, and diseases 25, 339 (1999)]. While aggregates are normally dissociated by moderate flow stresses,  under some pathological conditions the aggregation becomes irreversible, which leads to high blood viscosity and vessel occlusion. We perform here two-dimensional simulations to study the doublet dynamics under shear flow in different conditions and its impact on rheology. We sum up our results on the dynamics of doublet in a rich phase diagram in the parameter space (flow strength, adhesion energy) showing four different types of doublet configurations and dynamics. We  find that membrane tank-treading plays an important role in doublet disaggregation, in agreement with experiments on RBCs. A remarkable feature found here is that when a single cell performs tumbling (by increasing vesicle internal viscosity) the doublet formed due to adhesion (even very weak) remains stable even under a very strong shear rate. It is seen in this regime that an increase of shear rate induces an adaptation of the doublet conformation allowing the aggregate to resist cell-cell detachment. We show that the normalized effective viscosity of doublet suspension increases significantly  with the adhesion energy, a fact which should  affect blood perfusion in microcirculation.
\end{abstract}
\maketitle
\section{\label{sec:level1}Introduction}
The distribution of nutrients and oxygen to tissues and organs is ensured  by red blood cells (RBCs). Several cardiovascular dysfunctions and RBC anomalies may impair a proper blood perfusion to the organism.  For example, drepanocytosis \cite{di2016dense, lemonne2012increased}, which results in a stiff cytoplasm of RBCs under low oxygen conditions, compromises a proper RBC flow in small blood vessels due to a formation of occlusions. In several blood diseases, such as diabetes  and hypercholesterolemia, as well as coronary heart disease, an enhanced tendency of RBCs to form aggregates \cite{groeneveld1999relationship,prasad1993oxygen,  Lugovtsov2019,razavian1994influence,kaul2008sickle, potron1994fibrinogen, dunn2004fibrinogen, fusman2002red} has been reported. Fibrinogen (a protein contained in plasma) is the main cause of adhesion between RBCs \cite{brust2014plasma}. Under physiological conditions the range of human fibrinogen level is approximately 1.8 - 4 mg/ml \cite{comeglio1996blood}, and  RBCs aggregate and disaggregate reversibly under shear flow. In contrast, under pathological conditions the level of fibrinogen may become high enough leading to more stable aggregates \cite{razavian1992increase, dunn2004fibrinogen}. Foresto et al \cite{foresto2000evaluation} compared the aggregability of RBCs from diabetic and healthy patients, using direct microscopic observation and numerical processing. They found that the RBCs aggregation is highly enhanced in the case of diabetic patients compared to healthy  patients. It has been documented \cite{rogers1992decrease, mazzanti1997sialic} that within these patients  a degradation of the glycoprotein surface of the RBC membrane (which controls the electrical as well as steric repulsion between RBCs) favors aggregation.
In addition, it has been reported  that under pathological conditions, such as diabetes and drepanocytosis, the fibrinogen concentration is higher \cite{razavian1992increase, dunn2004fibrinogen, potron1994fibrinogen}. It is to be noted also that blood from 
 pregnant women shows an important aggregability level which is correlated with a high fibrinogen concentration \cite{huisman1988red}. It seems thus that enhanced adhesion can result both from an increase of fibrinogen level and alteration of RBCs surface properties (within diabetic patients, for example).

In a completely different context, that of microgravity during space missions, it has been reported that the analysis of blood from cosmonauts \cite{Markin1998} showed an increase of amylase activity. This enzyme is known to digest partially the glycocalyx on the surface of cells, such as RBCs and endothelial cells. This degradation promotes RBC-RBC aggregation \cite{Pot2011}  and their adhesion to macrophages. This highlights the potential  impact of long term space missions on cardiovascular dysfunctions.

Theoretically, two models are evoked to   describe the mechanism of adhesion between RBCs. The first one, which was prevailing for a long time, is the bridging model and has been adopted to account for fibrinogen and neutral dextran macromolecule-induced RBCs aggregation \cite{chien1973ultrastructural}. This model assumed that the proteins adsorb onto the RBC membrane and form a cross-link to the nearby RBC \cite{brooks1988mechanism}. The second model is the  depletion one, stating that  configurational entropy of the suspended molecules (e.g. fibrinogen) is lowered close to RBC surface, leading to a depletion layer, so that when the gap between two RBCs becomes of the order of depletion layer, the gap becomes less populated by fibrinogen molecules than elsewhere, and this results into osmotic attraction between RBCs.


The erythrocyte doublet aggregation is the primary process in the formation of erythrocyte aggregates in blood flow. Bertoluzzo et al \cite{bertoluzzo1999kinetic} studied the erythrocytes aggregation, using  light transmission through blood sample and observed that the aggregation starts by the formation of a collection of  RBC doublets. The doublet formation is a first basic building block for aggregation that should be clarified.
Ju et al \cite{ju2013effect} found numerically, using a Morse potential to account for cell-cell interaction (representing depletion forces), that RBC doublet with a homogeneous deformability sustains the adhesion, while an increased deformability difference between the two RBCs forming the doublet favors the doublet dissociation. In another numerical study,  Bagchi et al \cite{bagchi2005computational} modeled adhesion between RBCs by the ligand-receptor model (the bridging model). They analyzed the dependence of dynamics of RBCs aggregates on the adhesion energy. They found that the shearing force due to an imposed flow is more efficient to break the bonds than the normal pulling force.  Wang et al \cite{T.Wang} observed numerically that the doublet performs rotation  or undergoes  dissociation depending on the strength of inter-cellular force, the membrane deformability and the shear stress. Recently, Flormann et al \cite{flormann2017buckling} analyzed the doublet shape in a quiescent fluid {\it in vitro} and {\it in silico}, and observed that the contact surface of the doublet is  flat for weak adhesion and becomes of  sigmoid-type upon an increase of the adhesion energy (the protein concentration).



More recently Quaife et al. \cite{PhysRevFluids.4.103601} studied dynamics of a doublet under extensional and shear flows by using a 2D vesicle model. 
Under a linear shear flow (which is of interest to our study) they observed that the doublet undergoes a tumbling regime, that we shall refer to as  {\it  rolling} (for the sake of distinction with the classical single cell tumbling). By analyzing systematically dynamics of 2D vesicle doublets under various conditions, we find besides rolling,  three other distinct dynamics: flexible rolling (FR), rolling-sliding (RS) and flow alignment (FA). The FR motion corresponds to a situation where the two vesicles undergo global tumbling (rolling) but the contact interface between the two vesicles oscillates between flat and sigmoid shape. In the RS regime the two vesicles 
slide with respect to each other during rolling. The FA regime corresponds to the situation where the two vesicles align with the flow.
We present a general phase diagram of different phases of doublets. We shall also analyze the overall rheology. RBCs aggregation controls the rheological properties of blood which may constitute a promising diagnosis  for cardiovascular diseases \cite{dintenfass1974blood}.
\section{\label{sec:level2}Model and simulation method}
\subsection{\label{subsec:level1}Membrane model}
We consider a 2D model, namely phospholipid vesicles.
The two-dimensional vesicle model has proven to capture several features known for RBCs. Shapes like parachute and slipper \cite{guckenberger2018numerical, kaoui2009red}, dynamics, such as tumbling and tank-treading \cite{biben2011three, Kaoui2009VesiclesUS} are also manifested by both systems. We consider a set of phospholipid vesicles inside a straight channel, bounded by two rigid walls located at $y=0$ and $y = W$, where $W$ is the channel width. The vesicles are subject to a linear shear flow $v_{x}^{\infty}(y)= \dot{\gamma} y $ where $\dot{\gamma}$ is the shear rate. Periodic boundary conditions are used along $x$ axis (the flow direction). \medbreak
The force applied by the membrane on the surrounding fluid is obtained from the functional derivative of the following energy, which is the sum of three terms: the bending energy (Helfrich energy \cite{zhong1989bending}), the membrane incompressiblity contribution and the adhesion energy (Lennard-Jones potential)  between two vesicles:
\begin{equation}
E = \sum_{i}  E_{i}^{b} + \sum_{i \neq j} E_{i,j}^{adh}  ,
\end{equation}
where 
\begin{equation}
E_{i}^{b} = \frac{k}{2} \oint_{m_i} c^{2} ds + \oint_{m_i} \zeta ds
\end{equation}
is the bending and incompressibility energy of the $i$-th vesicle and 
\begin{equation} 
E_{i,j}^{adh}=\varepsilon \oint_{m_i}ds(\mathbf{X}_{i})\oint_{m_j}ds(\mathbf{X}_{j}) \phi(|\mathbf{X}_{i} -\mathbf{X}_{j}|)
\end{equation} 
is the energy of adhesion between $i$-th and $j$-th vesicle. The variable  $s$ represents the curvilinear coordinate on the vesicle contour, $c$ is the local curvature of the membrane, $k$ is the membrane bending rigidity, $\zeta$ is a local Lagrange multiplier associated with the constraint of local perimeter inextensibility, $\phi = -2\left(\dfrac{h}{r_{ij}}\right)^{6}+\left(\dfrac{h}{r_{ij}}\right)^{12}$ is Lennard-Jones potential which describes attractive interaction at long ranges and repulsive interaction at short ranges. Here $\mathbf{r}_{ij} = \mathbf{X}_{i}-\mathbf{X}_{j}$, where $\mathbf{X}_{i}$ and $\mathbf{X}_{j}$ are two position vectors of two material points on two different vesicles $i$ and $j$. $h$ is the equilibrium distance between two points of the $i$-th and $j$-th vesicle and $\varepsilon$ is the minimum energy associated to this distance. The functional derivative (providing the force) of the bending energy can be found in \cite{kaoui2008lateral}. The total force including vesicle-vesicle interaction has the following form :
\begin{equation}
\mathbf{f}(\mathbf{X}_{i}) = \mathbf{f}(\mathbf{X}_{i})^{b} + \mathbf{f}(\mathbf{X}_{i})^{adh}
\end{equation}
Where
\begin{equation}
\mathbf{f}(\mathbf{X}_{i})^{b}=\mathit{k}[\dfrac{d^{2}c}{d s^{2}}+\dfrac{ c^{3}}{2} ]\mathbf{n} -c \zeta \mathbf{n} + \dfrac{d \zeta}{d s} \mathbf{t}
\end{equation}
and 
\begin{equation}
\mathbf{f}(\mathbf{X}_{i})^{adh} = - \varepsilon \sum_{j \neq i} \int_{ m_{j}}\left[ \frac{d \phi(r_{ij})}{d r_{ij}} \left(\frac{\mathbf{r}_{ij}}{r_{ij}}\mathbf{.n}(\mathbf{X}_{i})\right)+c(\mathbf{X}_{i})\phi(r_{ij})\right]\mathbf{n}(\mathbf{X}_{i}) ds(\mathbf{X}_{j}).
\end{equation}
$\mathbf{n}$ and $\mathbf{t}$ are the normal and tangential unit vector respectively. The force can be rewritten in a dimensionless form:
\begin{equation}
\mathbf{\bar{f}}(\mathbf{X}_{i})^{b}=[\dfrac{d^{2}\bar{c}}{d \bar{s}^{2}}+\dfrac{ \bar{c}^{3}}{2} ] \mathbf{n} -\bar{c} \bar{\zeta} \mathbf{n} + \dfrac{d \bar{\zeta}}{d \bar{s}} \mathbf{t}
\label{force}
\end{equation}
and
\begin{equation}
\bar{\mathbf{f}}(\mathbf{X}_{i})^{adh} = - \bar{\varepsilon} \sum_{j \neq i}\int_{ m_{j}}\left[\frac{d \bar{\phi}(\bar{r}_{ij})}{d\bar{r}_{ij}
} \left(\frac{\mathbf{\bar{r}}_{ij}
}{\bar{r}_{ij}
}\mathbf{.n}(\mathbf{X}_i)\right)+\bar{c}(\mathbf{X}_i)\bar{\phi}(\bar{r}_{ij})\right]\mathbf{n}(\mathbf{X}_i) d\bar{s}(\mathbf{X}_{j})
\end{equation}
where dimensionless variables are defined as follows:
\begin{equation}
\bar{f} = \dfrac{R_{0}^{3} f}{\mathit{k}}, \enspace
\bar{\varepsilon} = \dfrac{R_{0}^{2} \varepsilon}{\mathit{k}}, \enspace
\bar{c} = c R_{0}, \enspace
\bar{s} = \dfrac{s}{R_{0}}, \enspace
\bar{r}_{ij}=\dfrac{r_{ij}}{R_0}, \enspace
\bar{\phi}(\bar{r}_ij)=\phi(\bar{r}_{ij}R_0).
\end{equation}
$R_{0} = \sqrt{ A / \pi }$ is the effective radius of the vesicle and $A$ is the enclosed vesicle area.

\begin{table}[hbtp]
\begin{ruledtabular}
\begin{tabular}{ l c c c c c c}
Fibrinogen concentration $mg/ml$ | & 0.898 & 2.391 & 4.197 & 5.402 & 6.597 & 8.098 \\
Interaction energy $\mu J/ m^{2}$ | & -1.884 & -2.719 & -3.748 & -4.655 & -4.922 & -6.566 \\
Dimensionless interaction energy | & 42.38 & 61.17 & 84.33 & 104.73 & 110.74 & 147.73 \\
\end{tabular}
\caption{\label{tab:1} Fibrinogen level versus Interaction energy between two RBC measured using atomic force microscopy \cite{brust2014plasma}}
\end{ruledtabular}
\end{table}

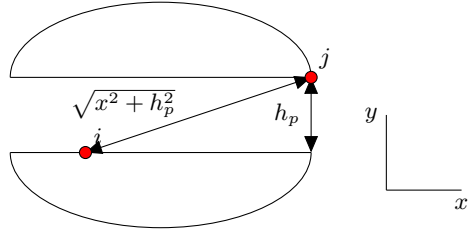
\begin{figure}[!htb]
\begin{tikzpicture}
\hspace{1.3cm}
\draw (0,0) arc(0:180:2cm and 1cm);
\draw (0,-1) arc(0:-180:2cm and 1cm);
\draw (-4,0) -- (0:0.01cm);
\draw (0,-1) (-4,-1) -- +(0:4cm);

\draw[>=triangle 45, <->] (-0.0,0) -- (-0.0,-1);
\draw[>=triangle 45, <->] (0,0) -- (-3,-1) ;

\foreach \Point/\PointLabel in {(0,0)/j, (-3,-1)/i}
\draw[fill=red] \Point circle (0.08) node[above right] {$\PointLabel$};

\node[text width=3cm] at  (1.01,-0.50) {$h_{p}$};
\node[text width=3cm] at  (-1.7,-0.35) {$\sqrt{x^{2} + h_{p}^{2}}$};

  \begin{scope}[y=1cm,xshift=1cm,yshift=-1.5cm] 
    \draw (0,0) -- (1,0) node[below] {$x$};
    \draw (0,0) -- (0,1) node[left] {$y$};
  \end{scope}
\end{tikzpicture}
\centering
\caption{\label{sketc} Notations}
\end{figure}

Previously, Brust et al \cite{brust2014plasma} quantified the interaction energy between two RBCs at various fibrinogen and dextran levels using single cell force microscopy. From their data (table~\ref{tab:1}), we can estimate which range of protein level corresponds to our simulation condition. When the contact length of two adhering vesicles is larger than the equilibrium distance $h$, the interaction energy per unit surface is practically the energy of adhesion between two infinite plates. The distance $h$ corresponds to the minimal energy between two points, and does not necessary correspond to the minimal energy for two planar surfaces. Let us denote that distance $h_p$ (see Fig.\ref{sketc}). We first calculate the energy for a given separation $h_p$ 
\begin{equation}
\varepsilon_{adh}=-\varepsilon \int \limits_{-\infty}^{\infty}\phi( \sqrt{x^{2}+h_p^{2}} ) dx=\varepsilon\int\limits_{-\infty}^{\infty}\left[2\left(\frac{h^2}{h_p^2+x^2}\right)^3-\left(\frac{h^2}{h_p^2+x^2}\right)^6\right]dx=\frac{3\pi h^6}{16h_p^5}-\frac{63\pi h^{12}}{256h_p^{11}},
\end{equation}
where $x$ is the coordinate along the vesicle-vesicle contact line.
We easily find that this energy has a minimum for $h_p =(231/320)^{1/6}h$, and is equal to 
\begin{equation}
\varepsilon_{adh} \simeq 1.6862 h \varepsilon.
\end{equation}
This will define the adhesion energy of the doublet. The dimensionless macroscopic  adhesion energy (to be referred to in all following results) is defined as 
\begin{equation}
\bar{\varepsilon}_{adh} =\dfrac{{\varepsilon_{adh}} R_{0}^{2}} {\mathit{k}}
\end{equation}
We will take typical values of RBC membrane rigidity $\mathit{k}= 4 \; 10^{-19} $ J and radius $R_0=3\; \mu$m.

\subsection{\label{subsec:level2}Boundary integral formulation}
The RBCs velocity and size are very small  corresponding to a small Reynolds number (in the range $10^{-4}$ to $10^{-2}$).  Here we consider the limit of  vanishing Reynolds number. In this case the fluid velocity inside and outside the vesicles is described by the Stokes equations:
\begin{equation}
 - \mathbf{ \nabla } p + \eta_{i} \Delta \mathbf{v} = 0
\end{equation}
\begin{equation}
\mathbf{ \nabla } . \mathbf{v} = 0,
\end{equation}
where $\eta_{i}$, with $i =0, 1$, is the viscosity of the internal (1) or the external (0) fluid, $p$ is the pressure and $\mathbf{v}$ is the velocity field.  
Due to the linearity of Stokes equations we can transform the set of fluid equations into an integral equation. 
 This is based on the use of Green’s function techniques\cite{pozrikidis1992boundary}, and  is a quite  accurate method for interface problems. More precisely, for a point $\mathbf{r_{0}}$ which belongs to a membrane, the velocity $\mathbf{v}(\mathbf{r_{0}})$ of that point has the following dimensionless expression:
\begin{widetext}
\begin{equation}
\mathbf{v}(\mathbf{r}_{0}) = \dfrac{2}{1 + \lambda} \mathbf{v}^{\infty}(\mathbf{r}_{0}) + \dfrac{1}{2 \pi C_{a} (1+\lambda)} \oint_{m} \underline{\underline{G}}(\mathbf{r}-\mathbf{r}_{0}).{\mathbf{f}}(\mathbf{r}) ds(\mathbf{r})+ \dfrac{(1-\lambda)}{2 \pi(1+\lambda)} \oint_{m} \mathbf{v}(\mathbf{r}).\underline{\underline{\underline{T}}}(\mathbf{r}-\mathbf{r}_{0}).\mathbf{n}(\mathbf{r}) ds(\mathbf{r}),
\label{integral}
\end{equation}
\end{widetext}
where $G_{ij}(\mathbf{r}-\mathbf{r_{0}})$ and $T_{ijk}(\mathbf{r}-\mathbf{r_{0}}) $ are the Green's functions corresponding to the two-dimensional channel bounded by two rigid walls \cite{thiebaud2013rheology} ($G_{ij}$ refers to the so-called single-layer contribution, while $T_{ijk}$ accounts for the double-layer contribution). The Green's functions satisfy the no-slip boundary condition at rigid walls.  $C_{a}$ is the capillary number and $\lambda$ the viscosity contrast, to be defined below.

We will analyze below the effect of the relevant blood flow and geometrical parameters on the rheological behavior of vesicle doublet and discuss the mechanism of separation. In order to preserve high accuracy we use Fourier basis  discretization of all functions and compute all derivatives in Fourier domain \cite{veerapaneni2009boundary,Dalal2020}. 
At high shear rates  numerical stability problems  may arise. In order to ensure long-term stability of the simulations, we keep carefully the vesicle perimeter and surface fixed. Normally, fluid incompressibility and membrane impermeability should keep the inner area of the vesicle  constant. However, a small drift due to numerical errors can not be
fully excluded. We compensate this drift by reinflating or deflating the  vesicle through homogeneous normal deformation.
\subsection{\label{subsec:level3}Dimensionless parameters}

Dimensionless numbers are used to describe the vesicle and the flow characteristics:
\begin{itemize}
\item The capillary number: allows to quantify the flow strength over bending rigidity of the membrane
\begin{equation}
C_{a} = \dfrac{\eta_{0} \dot{\gamma} R_{0}^{3}}{\mathit{k}} \equiv \dot{\gamma} \tau_{c}
\end{equation}
\item The confinement: describes the ratio between the effective diameter of the vesicle and the channel width
\begin{equation}
C_{n} = \dfrac{2 R_{0}}{W}
\end{equation}
\item The viscosity contrast: the ratio between the viscosities of the internal and external fluids
\begin{equation}
\lambda = \dfrac{\eta_{1}}{\eta_{0}}
\end{equation}
\item The reduced area: combining the vesicle perimeter  $L$  and its enclosed area $A$ 
\begin{equation}
\tau = \dfrac{(A/ \pi)}{(L/2 \pi)^{2}}
\end{equation}
\end{itemize}
This value will be set to 0.65 (inspired by that of human RBCs).
Throughout this paper, we will use the following scales: $R_{0}$ for the distance, $\tau_{c}$ for the time and $\eta_{0}$ for the viscosity.
\section{\label{sec:level3}Results}

The strategy followed in this work consists in preparing initially  two vesicles in the middle of the channel, separated by a small distance that allows them to adhere to each other in the absence of an applied flow. Once they adhere to each other and reach a steady state configuration, which depends on the adhesion strength, a shear flow is applied. The study of conformation of vesicle doublet in the absence of flow allowed us to perform  benchmarking of our code by reproducing previous results \cite{hoore2018effect, flormann2017buckling}. 

\subsection{\label{sec:level3-1}Effect of viscosity contrast, flow strength  and adhesion on the phase diagram of  doublet}
 We first analyzed the effect of the viscosity contrast $\lambda$, the capillary number $C_{a}$ and the adhesion energy $\varepsilon_{adh}$ on the dynamics of the doublet. The confinement parameter is set to $C_{n}=0.4$. We explored two values of viscosity contrast, $\lambda=1$ and $\lambda=10$. It is to be noted (for later purposes) that a  single vesicle exhibits a tank treading motion in the range from $\lambda =1.0$ (and also below that value)  to approximately $12$ (when $C_{n} = 0.4$ and $\tau=0.65$), beyond which it undergoes tumbling. This means that for both viscosity contrasts a single vesicle shows tank-treading. The critical value for transition from tank-treading to tumbling also depends on $C_n$. We will explore a less confined situation where a vesicle shows tumbling for $\lambda=10$. We will see in the next section that having tumbling will dramatically change our conclusion on the doublet separation process. 
\begin{figure}[hbtp]
\centering
\begin{subfigure}
\centering
\includegraphics[scale=0.40]{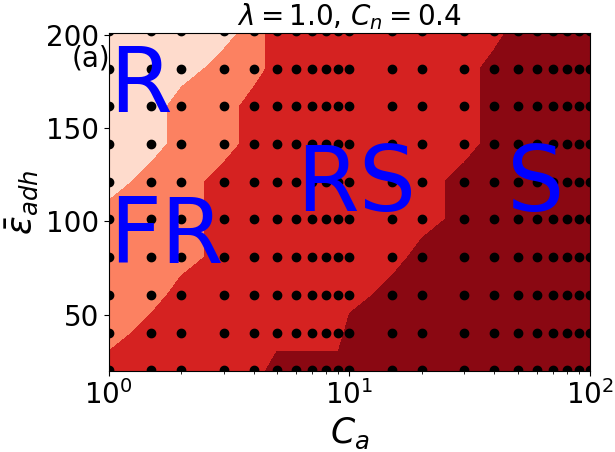}
\end{subfigure}
\hspace{0.1cm}
\begin{subfigure}
\centering
\includegraphics[scale=0.40]{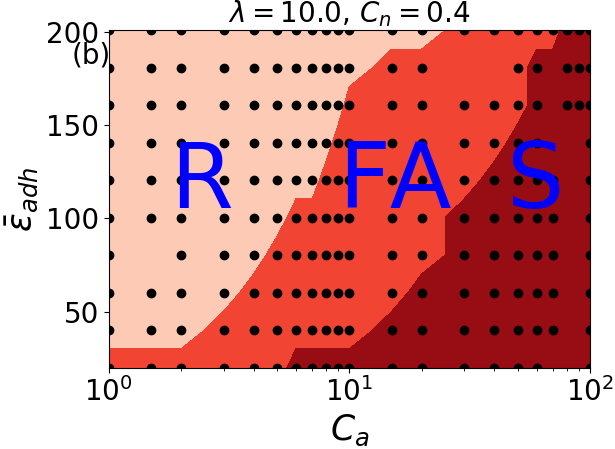}
\end{subfigure}
\caption{\label{fig:1} Phase diagrams showing the different behaviors of doublet in the parameter space  of the capillary number and the dimensionless macroscopic adhesion energy. The simulation data are shown as dots. (a) $\lambda = 1.0$, (b) $\lambda = 10.0$.}
\end{figure} 
 
\begin{figure}[hbtp]
\centering

\begin{subfigure}
\centering
\includegraphics[scale=0.35]{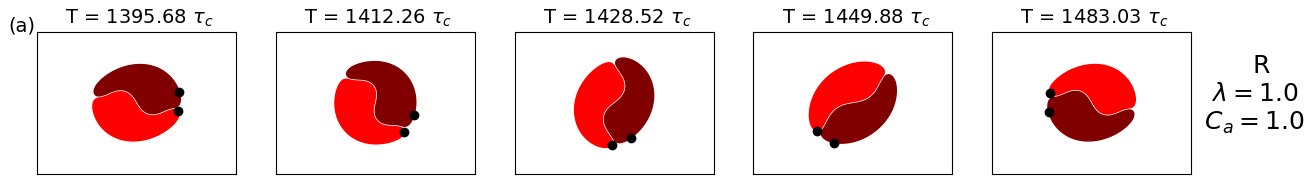}
\end{subfigure}

\begin{subfigure}
\centering
\includegraphics[scale=0.35]{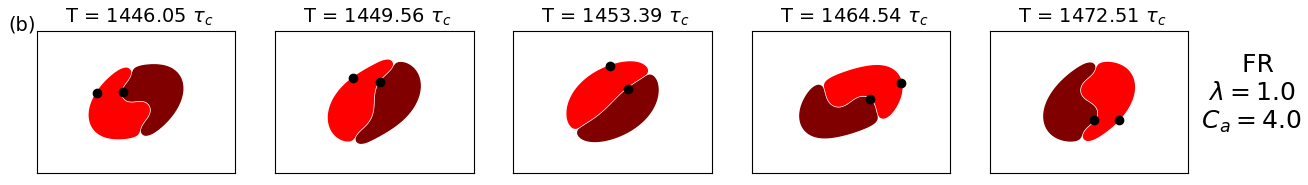}
\end{subfigure}

\begin{subfigure}
\centering
\includegraphics[scale=0.35]{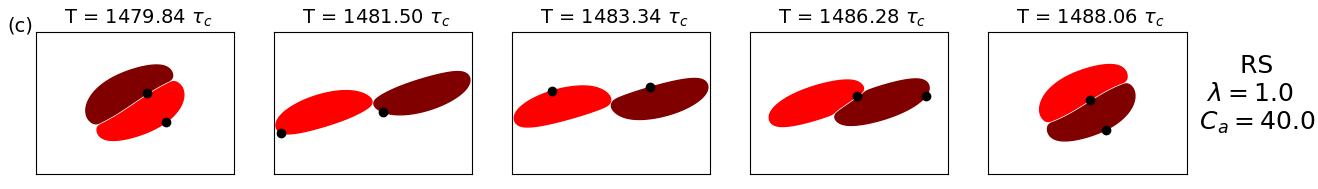}
\end{subfigure}

\begin{subfigure}
\centering
\includegraphics[scale=0.35]{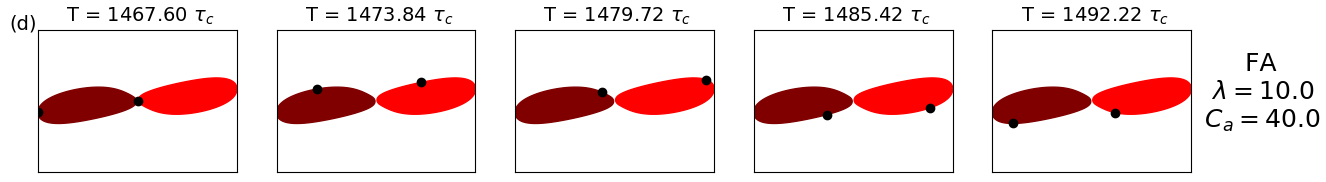}
\end{subfigure}
\caption{\label{fig:2} Snapshots showing the dynamics of vesicle doublet for different capillary number $C_{a}$ and viscosity contrast $\lambda$ . Here $\bar{\varepsilon}_{adh} = 202.00$ and $C_{n} = 0.4$. The snapshots are taken over one period of the doublet dynamic.}
\end{figure}

 Once the doublet reaches a steady-state configuration at equilibrium (i.e. in the absence of external flow), we applied a shear flow with different values of shear rate and analyzed the dynamics exhibited by the doublet. We found a quite rich phase diagram (Fig.~\ref{fig:1}-a) in a wide range of capillary number $C_{a}$ and dimensionless macroscopic adhesion energy $\bar{\varepsilon}_{adh}$. Three regimes have been identified in the case of $\lambda = 1$: (i) at low capillary number the rolling phase: the two vesicles remain attached with constant contact length and show a rolling motion (akin to  tumbling; see Fig.~\ref{fig:2}-a). This regime is denoted as $R$ in Fig.~\ref{fig:1}-a.  At large enough adhesion strength, the contact line between the two vesicles becomes  of sigmoid type (as seen in Fig.~\ref{fig:2}-a). (ii) When the adhesion energy decreases the contact interface shape transits from sigmoid to a flat interface during time as shown in the snapshots of (Fig.~\ref{fig:2}-b)). These two states (flat and sigmoid interface) were identified in a previous study \cite{flormann2017buckling} as two equilibrium states (in the absence of external flow), and it was shown that upon increasing adhesion strength there is (a supercritical) bifurcation from the flat towards a sigmoid state; the transition from flat to sigmoid interface is a symmetry breaking bifurcation because the mirror symmetry with respect to a plane orthogonal to the interface is lost for the sigmoid phase. Here this symmetry is already broken by the shear flow so the boundary is never exactly flat. However, we can visually observe an oscillating solution in time between a quasi-flat (sigmoid with a small amplitude) state and a developed sigmoid solution. It has seemed to us useful to distinguish this motion as a different  "state"  from $R$ phase, and we refer to it as $FR$  (flexible rolling) phase.  The demarkation line between $R$ and $FR$ is set such that when the amplitude of oscillation of the  distance (in unit of $R_0$) separating mass centers of the two vesicles  reaches  5$\%$ (in $R$ phase the amplitude is less than $5\%$, and it is larger in $FR$ phase).  
 A close inspection shows that the pitchfork bifurcation from flat to sigmoid in the absence of flow \cite{flormann2017buckling}  becomes an imperfect bifurcation in the presence of flow.
 (iii)   Increasing the capillary number the phase  ($FR$)  undergoes a transition towards a phase of rolling+sliding ($RS$; the doublet shows rolling accompanied with  a sliding between the vesicles; the length contact changes during time by oscillating between two values (Fig.~\ref{fig:2}-c)). The sliding of the two vesicles on each other is due to the competition between the aggregation and dis-aggregation (flow) forces, as well as  due to  tank treading of each membrane.  This regime is denoted as $RS$ in (Fig.~\ref{fig:1}-a).  (iv) At high capillary number the separation between the vesicles takes place, meaning the dis-aggregation force due to flow is high enough to overcome the adhesion between vesicles. This regime is denoted as $S$ in Fig.~\ref{fig:1}-a. We will see below situations where a doublet may persist whatever the magnitude of the shear flow is.

\begin{figure}[hbtp]
\centering
\includegraphics[scale=0.48]{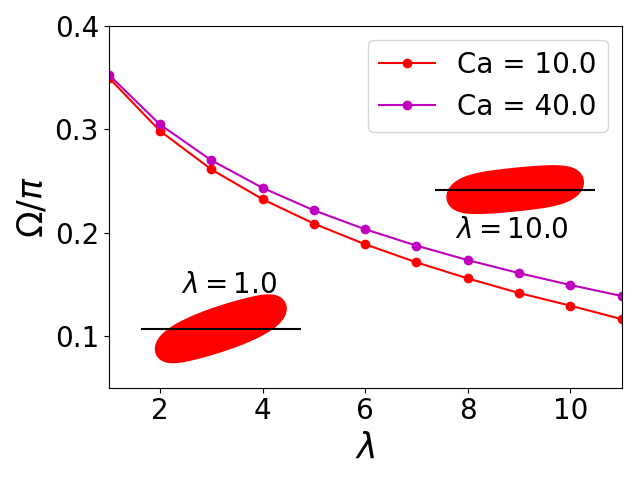}
\caption{\label{fig:4}  Inclination angle of a single vesicle as a function of viscosity contrast $\lambda$. The angle decreases with $\lambda$. The snapshots show the vesicle under shear flow of $C_{a} = 10.0$.}
\end{figure}

Subsequently we have evaluated the effect of the viscosity contrast on the phase diagram discussed above. The results are shown in Fig.~\ref{fig:1}-b for $\lambda=10$. Several observations are made. First, the rolling region becomes wider if viscosity contrast is increased. Second, the rolling-sliding phase is absent for this viscosity contrast, in favor of a flow-alignment (FA) phase (Fig.~\ref{fig:2}-d)).  In this phase the vesicles align with the flow direction,  remain attached and they show a tank treading motion of their membrane. 
  In order to shed some light on the origin of the flow alignment at high viscosity contrast, we performed simulations on a single vesicle at fixed values of the capillary number $C_{a}$, the confinement $C_{n}$ and we varied only the viscosity contrast $\lambda$. Figure ~\ref{fig:4} shows that the inclination angle $\Omega$ of a single vesicle (the angle between the long axis of the vesicle and the flow direction) decreases with the viscosity contrast, until it aligns with the flow direction at high viscosity contrast. In this configuration (flow alignment) the doublet is in an orientation with a small extensional tension which is not efficient to enforce the two vesicles to slide with respect to each other.  

\subsection{\label{sec:level3-2}The mechanism of doublet separation}
One major question is whether or not doublet formation is reversible {\it in vivo}, and whether there is any simple  criterion (or hint)  to answer this question. With this regard, 
Chien et al \cite{chien1990dynamics} investigated experimentally the dis-aggregation of RBC doublets under oscillatory shear flow. They prepared two adhered cells in a flow channel where  the bottom cell adheres to a fixed plane and a polystyrene latex particle is used as a marker on the top cell. By applying an oscillatory shear flow they observed that the velocity of the latex bead is twice the velocity of the upper cell. From this it was concluded that the detachment occurs as rolling of the top cell along the bottom one and not as sliding. In other words, the bottom surface of the top cell remained stationary in the lab frame except at the point where the detachment occurred, while the velocity of the top surface of the top cell was twice that of its center of mass. This motion is similar to tank-treading in the reference frame co-moving with the top cell, where its top and bottom surfaces move with opposite velocities, while the overall shape of the cell remains virtually unchanged. Thus the ability of the cell membrane to tank-tread is essential for the doublet separation. We now use our model to see whether preventing the membrane from tank-treading would have an effect on doublet dissociation.
\begin{figure}[hbtp]
\centering
\begin{subfigure}
\centering
\includegraphics[scale=0.40]{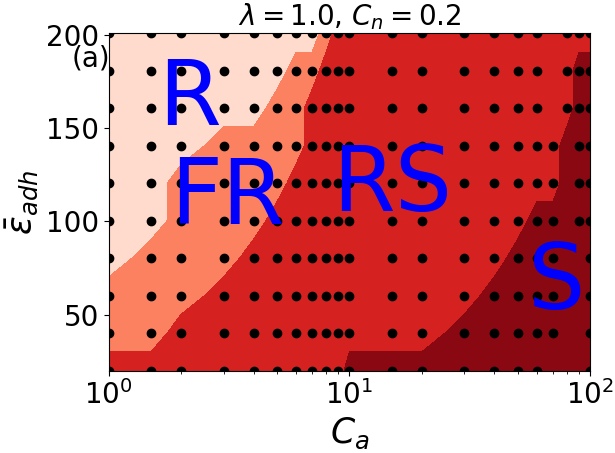}
\end{subfigure}
\hspace{0.1cm}
\begin{subfigure}
\centering
\includegraphics[scale=0.40]{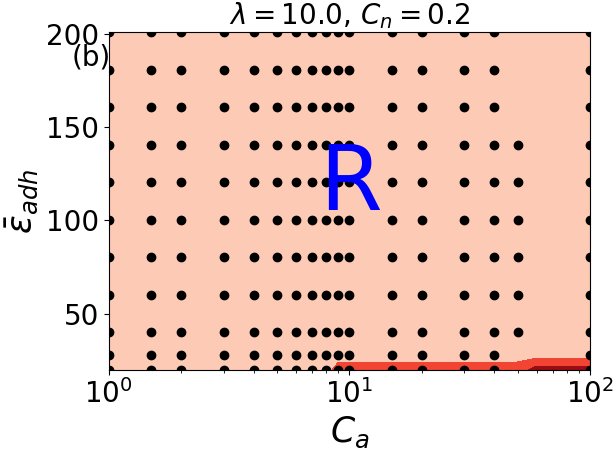}
\end{subfigure}
\caption{\label{fig:5} Phase diagrams showing the different behaviors of doublets in the  parameter space of the capillary number and the dimensionless macroscopic adhesion energy. The simulation data are shown as dots. (a) $\lambda = 1.0$, (b) $\lambda = 10.0$.}
\end{figure}

\begin{figure}[hbtp]
\centering
\includegraphics[scale=0.39]{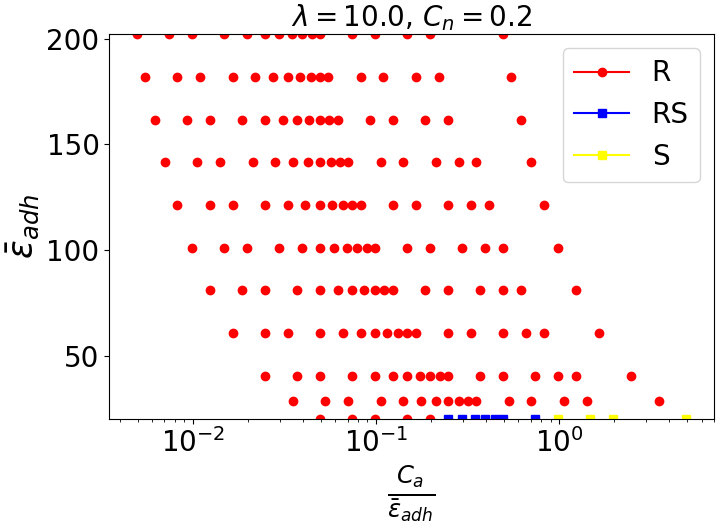}
\caption{\label{scaling} Phase diagrams showing the different behaviors of doublets in the  parameter space of the dimensionless macroscopic adhesion energy and the ratio of capillary number over dimensionless macroscopic adhesion energy. The simulation data are shown as dots.}
\end{figure}

\begin{figure}[hbtp]
\centering
\includegraphics[scale=0.5]{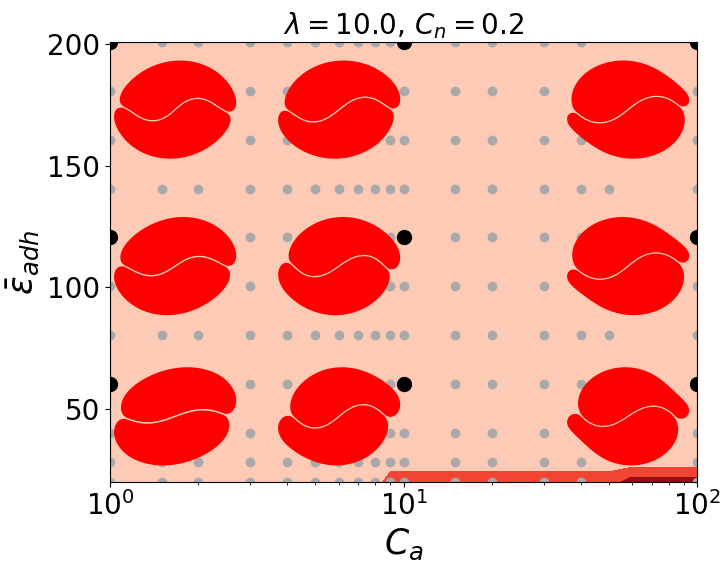}
\caption{\label{adapt} Phase diagrams (same as in Fig.\ref{fig:1}) showing few  shapes  of doublet in the  parameter space of the capillary number and the dimensionless macroscopic adhesion energy. The simulation data are shown as dots. The darkest spots indicate the pairs of parameters for which the shape is shown. We clearly see shape adaptation as $C_a$ increases.}
\end{figure}

We have seen above that there exists (Fig.~\ref{fig:1}) a region of separation (region $S$) for a given  set of  parameters. In both diagrams of Fig.~\ref{fig:1} the separation phase (S) is preceded either by  $RS$ or $FA$ phase. In both of the latter two phases the  membrane undergoes tank-treading. The question naturally arises of  whether or not the separation phase is associated with the existence of membrane tank-treading. We have thus investigated if the suppression (or a significant reduction) of membrane tank-treading may affect separation. For that purpose we have chosen a wide enough channel ($C_{n} = 0.2$) and only varied  the viscosity contrast $\lambda$ in order to reduce membrane tank-treading. As a guide, we  have first analyzed  the case of a single vesicle and determined the critical $\lambda$ for the transition from tank-treading to tumbling, and found $\lambda \simeq 7.0$ (note that  when $C_n=0.4$  this transition takes place at about $\lambda =12$). This means that for $C_n=0.2$ a single vesicle  shows tank-treading  for $\lambda=1$ and tumbling for $\lambda =10$. We have analyzed the phase diagram for these two values and the results are shown in Fig.~\ref{fig:5}. For $\lambda=1$ the same overall picture is found as in Fig. Fig.~\ref{fig:1}a. However, the situation is drastically different when $\lambda = 10.0$ (Fig.~\ref{fig:5}-b). Indeed,  we see that $RS$ and $S$  phases are almost absent, and they appear only for extremely weak adhesion, far below physiological ranges.  In other words, the doublet seems to be very robust even for a very large capillary number. We have attempted to dig further into this result. 
  The  doublet precise shape is a compromise between adhesion, which tends to increase vesicle-vesicle interface (like sigmoid shape) and bending energy, which tends to minimize deformation, favoring a flatter interface. Deformation ability can be measured by $C_a$, and dimensionless macroscopic adhesion energy by $\bar{\varepsilon}_{adh}$. An enlightening representation of our result   Fig.~\ref{fig:5} is to plot $\bar{\varepsilon}_{adh}$ as a function of $C_a/\bar{\varepsilon}_{adh}$. The results are shown in Fig. \ref{scaling}. This clearly shows that the shape adapts itself to shear flow. Indeed, if that were not the case, namely that the doublet conformation were independent of $C_a$ for given adhesion energy, then the R/RS and RS/S phase borders observed for the lowest dimensionless macroscopic adhesion energies would continue vertically to high dimensionless macroscopic adhesion energies  (in the representation of  Fig. \ref{scaling}). In other words, the transition value of $C_a/\bar{\varepsilon}_{adh}$ would be independent of adhesion energy. Indeed,  if bending energy saturates,  the only  remaining two energy scales are shear and adhesion ones, so that the critical $C_a/\bar{\varepsilon}_{adh}$  would have been a numerical constant. The fact that the phase diagram in the new representation shows the same trend as in    Fig.~\ref{fig:5}-b is a clear indication that the doublet adapts its shape in order to escape dissociation.  Figure \ref{adapt} shows some doublet shapes highlighting adaptation to shear flow. 
Our results are consistent with the work of Chien et al. \cite{chien1990dynamics}, in that membrane tank-treading plays an important role in doublet dissociation. A single RBC the possibility of membrane tank-treading depends on capillary number, defined by $C_s=\eta \dot \gamma /G$, where  $G$ is the cytoskeleton shear modulus (typically $G\simeq \; 4 \mu N/m$).  For low enough $C_s$ tumbling prevails whereas membrane tank-treading is possible at high $C_s$ \cite{Yazdani2011,Fischer2013}. Taking for $\eta$ the value of water viscosity (which is close to the  plasma one) and $R_0\sim 3 \; \mu m$, we obtain $C_s\sim 10^{-3} \dot \gamma$ (with $\dot\gamma$ in unit of $s^{-1}$). For healthy RBCs the transition between tumbling and   tank-treading takes place at about   $C_s\sim 0.1$ \cite{Yazdani2011,Fischer2013}. In human vascular networks  we expect membrane tank-treading to take place in arterioles only (the only vasculature site where shear rate can reach values of about few $10^3$ s$^{-1}$). In some RBCs diseases, such as                thalassemia \cite{advani1992oxidative}, sickle-cell disease \cite{brandao2003optical} and malaria \cite{glenister2002contribution}, the membrane shear modulus as well as cytoplasm viscosity may be significantly higher than healthy ones. The corresponding shear rate beyond which membrane tank-treads may become significantly larger for pathological cells \cite{Yazdani2011} that it occurrence in vivo becomes unlikely. We can speculate that in this case RBC doublets and larger aggregates become irreversible, compromising thus a proper blood perfusion to tissues and organs.

\subsection{\label{sec:level3-3}The effect of adhesion energy on the instantaneous  normalized viscosity.}
The goal of the next two paragraphs is to explore the effect of doublet dynamics discussed above on the rheological behaviors of a doublet suspension. In order to quantify rheology we analyze the normalized viscosity. Here we consider the very dilute regime. The effective viscosity can be written in the following form:
\begin{equation}
\eta = \eta_{0} ( 1 + [ \eta ] \varphi)
\end{equation}
where $\varphi$ is the vesicle   concentration, and $[\eta]$ is the normalized viscosity (called also the intrinsic viscosity),  representing the doublet contribution to the viscosity.
The effective viscosity is the ratio of the $xy$ component of stress tensor to the applied shear rate:
\begin{equation}
\eta = \frac{<\sigma_{xy}>}{\dot{\gamma}}
\end{equation}
Where bracket $<...>$ means a surface average (i.e. average over the simulation area). Following Batchelor \cite{batchelor1970stress}, the normalized viscosity is given by:
\begin{equation}
[\eta] = \frac{\eta - \eta_{0}}{\eta_{0} \varphi} = \frac{1}{\eta_{0} A \dot{\gamma}} \sum_i\left[ \int_{m_i} y f_{x} ds + \eta_{0} (\lambda - 1) \int_{m_i} (n_{x} v_{y} + n_{y} v_{x})ds \right] 
\end{equation}
 The first term of the normalized viscosity describes the dynamical contribution which is due to the membranes force, whereas the second term represents  the kinematic contribution of the vesicle (the membranes velocity). 
 
\begin{figure}[hbtp]
\centering
\begin{subfigure}
\centering
\includegraphics[scale=0.315]{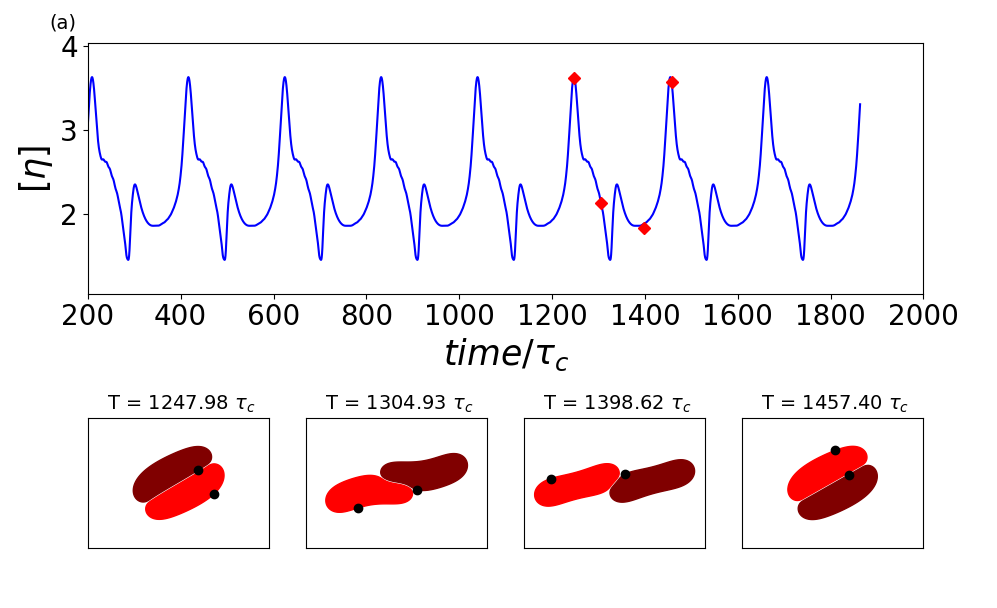}
\end{subfigure}
\begin{subfigure}
\centering
\includegraphics[scale=0.315]{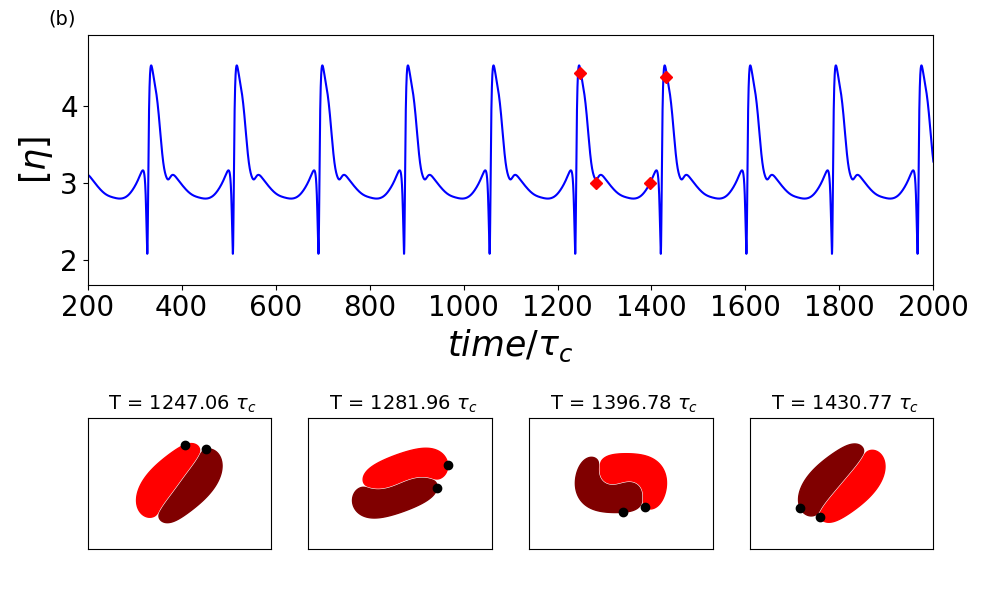}
\end{subfigure}
\hspace{0.05cm}
\begin{subfigure}
\centering
\includegraphics[scale=0.315]{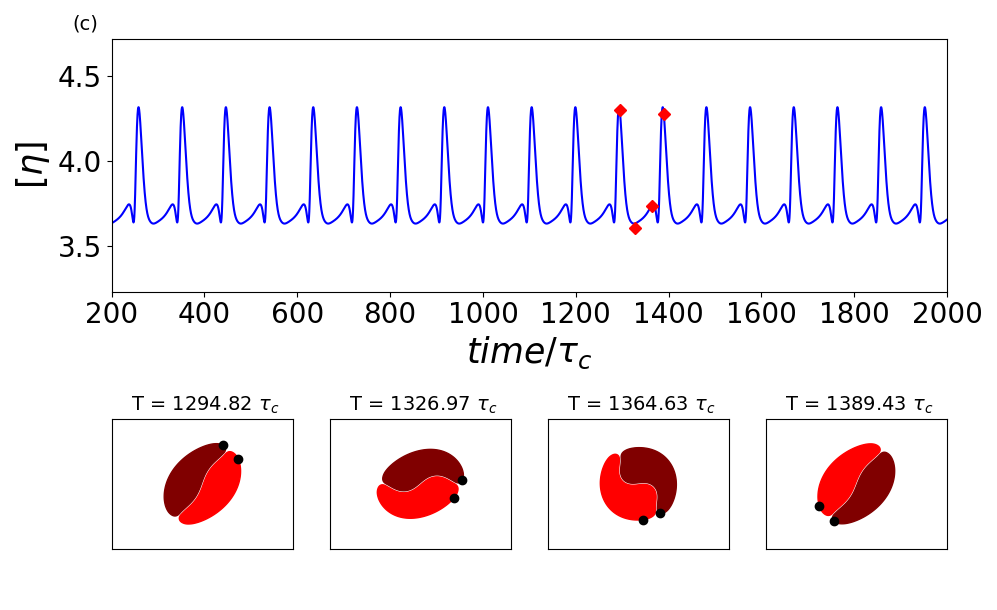}
\end{subfigure}
\begin{subfigure}
\centering
\includegraphics[scale=0.315]{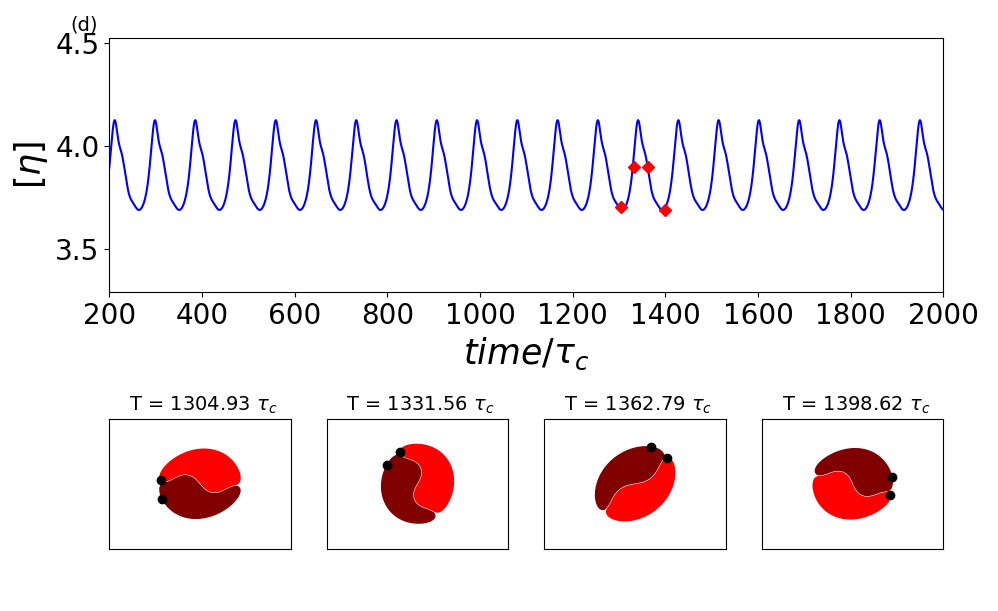}
\end{subfigure}
\caption{\label{fig:6} Evolution of the normalized viscosity $[\eta]$ as function of time. The snapshots are taken over one time period as shown with the red dots. Here $\lambda = 1.0$, $C_{n} = 0.4$ and $C_{a} = 1.0$ (a) $\bar{\varepsilon}_{adh} = 20.20$, (b) $\bar{\varepsilon}_{adh} = 40.40 $, (c) $\bar{\varepsilon}_{adh} = 80.83 $, (d) $\bar{\varepsilon}_{adh} = 202.00 $}
\end{figure}

We first analyzed how the normalized viscosity $[\eta]$ changes during time for increasing adhesion energy with a viscosity contrast $\lambda = 1.0$ and low capillary number $C_{a} = 1.0$. Figures~\ref{fig:6}a-d show that the viscosity of the suspension is periodic with time.  At very low adhesion strength the doublet shows $RS$ regime (Fig.~\ref{fig:6}a).  This state undergoes a transition towards the $R$-regime at high adhesion strength ((Fig.~\ref{fig:6}d). This transition is accompanied by an increase of the amplitude of the normalized viscosity. In the $R$-phase the contact interface of the doublet does not evolve with time. The amplitude as well as the period of the normalized viscosity oscillation decreases with the adhesion strength. This is attributed to the fact that the two vesicles become more and more pinned to each other (Fig.~\ref{fig:6}d) as the adhesion energy increases, so that the overall  cross section of the doublet which is exposed to the flow decreases, opposing thus less resistance.

\subsection{\label{sec:level3-4} Time average rheology of doublet suspension}
\begin{figure}[hbtp]
\centering
\begin{subfigure}
\centering
\includegraphics[scale=0.35]{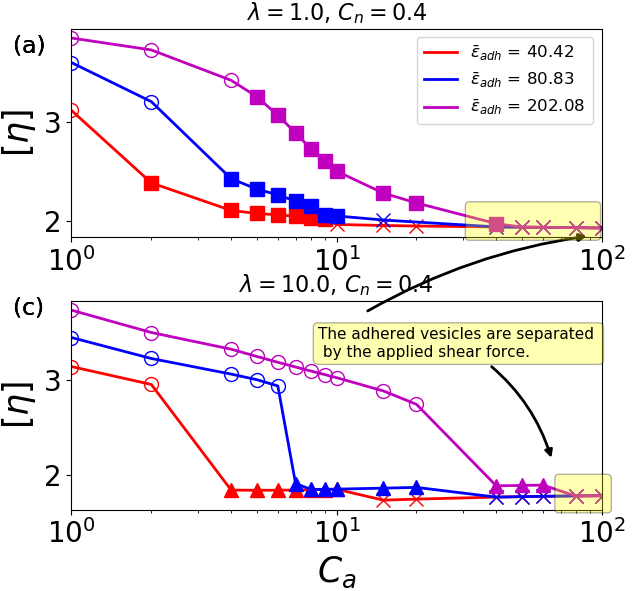}
\end{subfigure}
\hspace{0.1cm}
\begin{subfigure}
\centering
\includegraphics[scale=0.35]{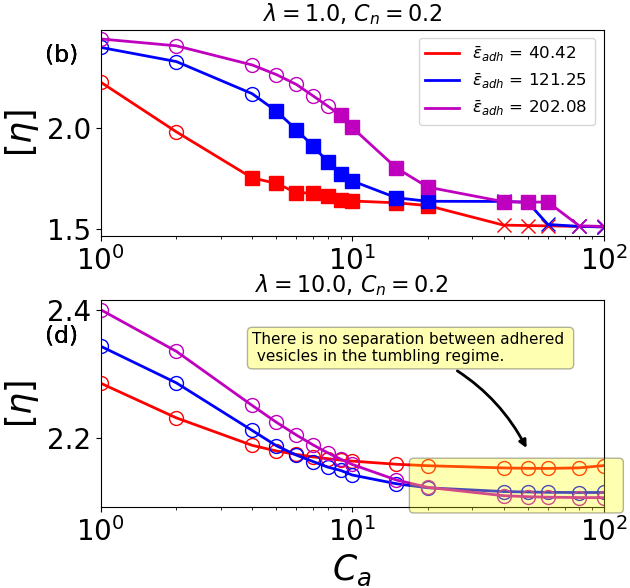}
\end{subfigure}
\caption{\label{fig:7} The normalized viscosity $[\eta]$ as a function of capillary number $C_{a}$ for different dimensionless macroscopic adhesion energy $\bar{\varepsilon}_{adh}$. The simulation data are shown as dots. (a) $\lambda = 1.0$ and $C_{n} = 0.4$, (b) $\lambda = 10.0$ and $C_{n} = 0.4$, (c) $\lambda = 1.0$ and $C_{n} = 0.2$, (d) $\lambda = 10.0$ and $C_{n} = 0.2$,}
\end{figure}

As seen above a single vesicle is known to exhibit both tank-treading motion (at low viscosity contrast) and tumbling motion (at high viscosity contrast). A suspension of vesicles 
 exhibit both shear thinning and shear thickening depending on viscosity contrast\cite{ouhra2018shear}. Here we find that the doublet  suspension always exhibits shear thinning for the set of parameters explored so far. The results are shown in Fig.~\ref{fig:7} (note that capillary number axis in Fig.~\ref{fig:7} is shown in logarithmic scale). In all the three studied cases (Fig.~\ref{fig:7}a,b,c) the viscosity collapses by about 50$\%$. A common feature shown in Fig.~\ref{fig:7}a,b,c  is that all curves (obtained for different adhesion energies) collapse on the same curve for high enough capillary number. This collapse corresponds to the situation where all doublets are dissociated. Figure~\ref{fig:7}d does not show the same behavior, in that the curves do not collapse at large capillary number, due to the absence of dissociation. This figure shows a peculiar behavior: at low capillary number the suspension with high dimensionless macroscopic adhesion energy has a higher normalized viscosity (which is quite intuitive), but at higher capillary number the opposite is found. 

Let us  provide an argument for this behavior. 
For a low dimensionless macroscopic adhesion energy(say the red curve in Fig.~\ref{fig:7}d)  and at low capillary number the doublet shows the $R$ motion where the contact length between the two vesicles oscillates in time while keeping sigmoid shape   (not to be confused with $FR$; see Movie 1 in the supplemental material \cite{SM}). For a higher adhesion (say the blue curve in Fig.~\ref{fig:7}d), and at low capillary number, the contact surface  has a persistent sigmoid shape. In other words for a high enough adhesion the doublet is quite rigid in its configuration yielding a higher viscosity, as intuitively expected. When the capillary number increases, for the low adhesion case (the red curve in Fig.~\ref{fig:7}d), the shear stress is strong enough to pull on the doublet leading to a peeling off the tail of each vesicle at the poles (see Movie 3 in the supplemental material \cite{SM}). The extra tails that stick out of the vesicles increase the cross section of the doublet, leading to a higher viscosity than the case where adhesion is stronger (preventing tails to stick out of the doublet, see Movie 4 in the supplemental material \cite{SM}). This explains why, at high capillary number, for weak adhesion the viscosity is higher than for strong adhesion (the blue and red curves intersect).
   

\begin{figure}[hbtp]
\centering
\begin{subfigure}
\centering
\includegraphics[scale=0.48]{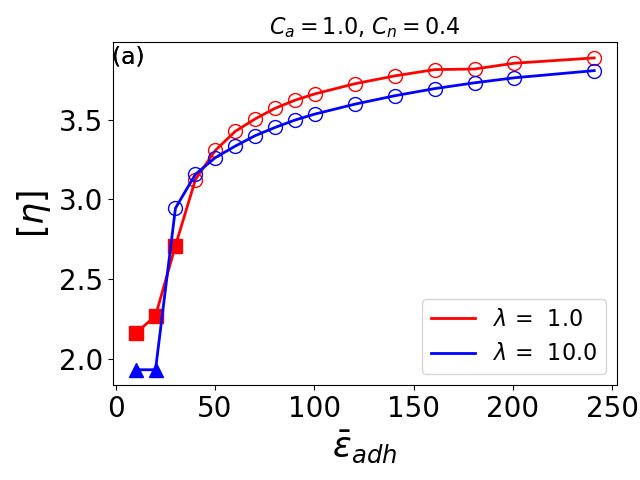}
\end{subfigure}
\hspace{0.1cm}
\begin{subfigure}
\centering
\includegraphics[scale=0.48]{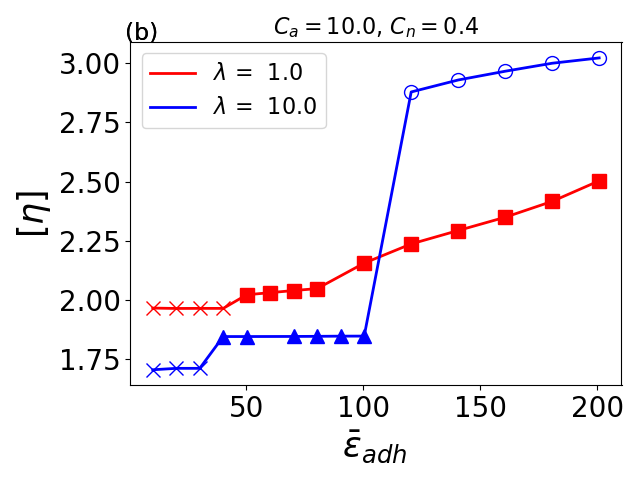}
\end{subfigure}

\begin{subfigure}
\centering
\includegraphics[scale=0.48]{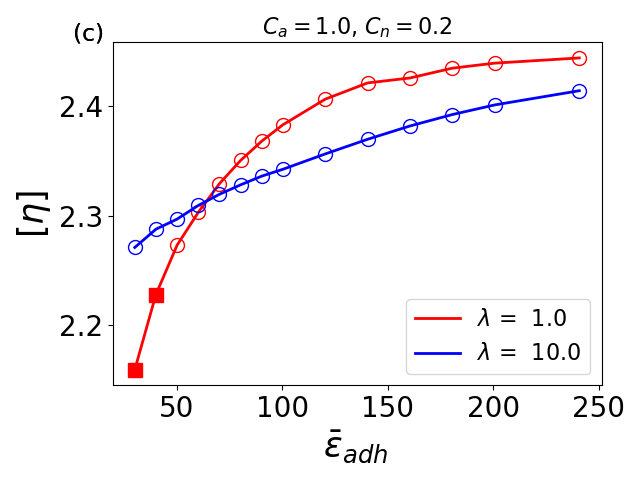}
\end{subfigure}
\hspace{0.1cm}
\begin{subfigure}
\centering
\includegraphics[scale=0.48]{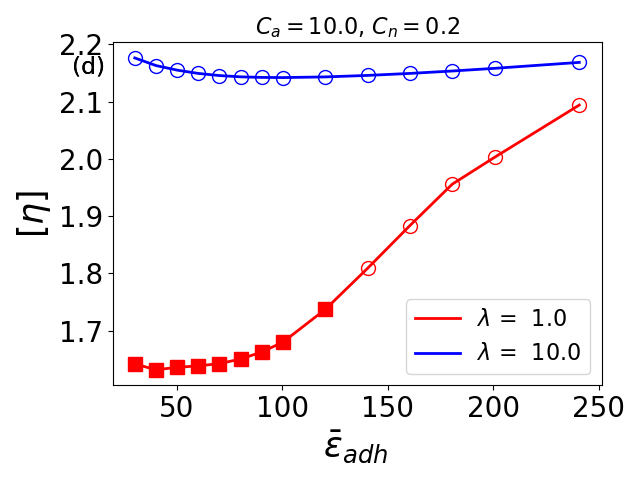}
\end{subfigure}
\caption{\label{fig:8} The normalized viscosity $[\eta]$ as a function of the dimensionless macroscopic adhesion energy $\bar{\varepsilon}_{adh}$ for different values of viscosity contrast and confinement. The simulation data are shown as dots. (a) $C_{a} = 1.0$ and $C_{n} = 0.4$, (b) $C_{a} = 100.0$ and $C_{n} = 0.4$, (c) $C_{a} = 1.0$ and $C_{n} = 0.2$, (d) $C_{a} = 100.0$ and $C_{n} = 0.2$. Circle symbol represents the rolling phase, square: rolling + sliding, triangle: flow alignment, cross: separation.}
\end{figure}

Finally, let us quantify the effect of the dimensionless macroscopic adhesion energy on the normalized viscosity. Note that the range of the dimensionless macroscopic adhesion energy from $5.0$ to $90.0$ corresponds to the physiological conditions of fibrinogen level (these values are estimates from \cite{brust2014plasma}). Pathological conditions correspond to a  value of dimensionless macroscopic adhesion energy greater than $90.0$. Figure ~\ref{fig:7} shows that  $[\eta]$ increases monotonically with $\bar{\varepsilon}_{adh}$, except for the case of Fig.~\ref{fig:8}d with $\lambda=10$  which has been discussed Fig.~\ref{fig:7}d where we have seen an inversion of normalized viscosity behavior as a function of the dimensionless macroscopic adhesion energy. Note also that the range of variation of normalized viscosity can be ample enough (it can attain a factor of two; see Fig.~\ref{fig:8}a).

\section{\label{sec:level4}Summary and concluding remarks}
To summarize, in this paper we employed a boundary integral method  and vesicle model to study the dynamics and the rheology of vesicle doublet under shear flow. We varied the dimensionless macroscopic adhesion energy in this study from $5.0$ to $250.0$, a range  which corresponds to  physiological and pathological conditions.  We found that the doublet can exhibit rolling, flexible rolling, rolling+sliding, aligned vesicles and that separation depends on several parameters (capillary number, adhesion energy and viscosity contrast). A remarkable feature is that when each single cell exhibits  tumbling (for example due to a high enough internal viscosity),  the doublet becomes quite stable even for an extremely large shear stress (the separation region is almost absent in the phase diagram, Fig.\ref{fig:5}). Indeed, the doublet adapts  its spatial configuration to the applied flow in a way to escape dissociation. In vivo RBC membrane  may perform tank-treading in arterioles only. However, several RBCs  pathologies are associated with  increased membrane rigidity or cytosol viscosity. This  may result in a collapse of the membrane tank-treading ability (even in arterioles) causing a stability of doublets, and impairing blood perfusion in microcirculation.

The rheological study showed shear thinning, and a quite significant increase of viscosity with adhesion energy. Rheology may be an interesting alternative for a systematic blood diagnosis and estimate of adhesion energy. A systematic numerical study in 3D  including cytoskeleton is necessary before drawing more applicative conclusions.


\begin{acknowledgments}
We acknowledge financial support from CNES (Centre National d'Etudes Spatiales), and the French-German University Programme "Living Fluids" (Grant CFDA-Q1-14). The simulations were performed on the Cactus cluster of the CIMENT infrastructure, which is supported by the Rh\^{o}ne-Alpes region (Grant No.CPER07\_13 CIRA).
\end{acknowledgments}
\newpage
\bibliographystyle{abbrv}
%
\end{document}